\documentstyle[12pt,fullpage]{article}
\begin{document}

\vspace*{2cm}
\begin{center}
{\large\bf Mean-field theory for a spin-glass model of neural 
networks:\ TAP free energy and paramagnetic to spin-glass transition} 

\vspace{1cm}
K Nakanishi\dag\footnote{nakanish@kanazawa-gu.ac.jp} \  
and H Takayama\ddag\footnote{takayama@issp.u-tokyo.ac.jp} \\

\vspace{5mm}
\dag Faculty of Business Administration and Information Science, \\
Kanazawa Gakuin University, Kanazawa 920-13, Japan \\
\ddag Institute for Solid State Physics, University of Tokyo, \\
Roppongi, Minato-ku, Tokyo 106, Japan

\vspace*{2cm}
\end{center}

\vspace*{1cm}
\begin{abstract}
An approach is proposed to the Hopfield model where the mean-field treatment 
is made for a given set of stored patterns (sample) and then the statistical 
average over samples is taken.  This corresponds to the approach made by 
Thouless, Anderson and Palmer (TAP) to the infinite-range model of spin 
glasses.  Taking into account the fact that in the Hopfield model there exist 
correlations between different elements of the interaction matrix, we obtain 
its TAP free energy explicitly, which consists of a series of terms exhibiting 
the cluster effect. Nature of the spin-glass transition in the model is 
also examined and compared with those given by the replica method as well 
as the cavity method.
\end{abstract}

\clearpage
\vspace*{-3mm} 
\section{Introduction}

Neural networks are systems in which a great number of neurons are connected 
with each other by synapses.  The neurons basically take the two states, i.e., 
the firing and non-firing states.  A neuron is firing if stimuli coming 
from (thousands of) neighboring neurons exceed a threshold. The neuron thus 
firing in turn affects neighboring neurons.  These features are reminiscent of 
an Ising spin system with long ranged interactions.

Hopfield [1] pointed out that the neural networks can be described by a
mathematically equivalent model to that of spin glasses if couplings through
synapses are symmetric and random.  This suggests that the various methods 
developed  for spin glasses are applicable to the neural networks.  
Indeed, numbers of studies have been made on this model since then [2-5].
Among others, the work made by Amit, Gutfreund and Sompolinsky (AGS) [5] 
is worth noticing.  Applying the replica method, which is a mathematical 
trick to calculate the free energy, they investigated the Hopfield model to 
find that it exhibits a feature of the associative memory in a certain region 
in the $T-\alpha$ plane, where $T$ is temperature and $\alpha=p/N$ is the 
ratio of the number of stored patterns $p$ to that of neurons $N$. 
The region is called the retrieval feromagnetic (FM) phase. It was also shown 
by AGS that the model has another ordered phase, called the spin-glass (SG) 
phase, besides a disordered paramagnetic (PM) phase at highest temperatures.

Sherrington and Kirkpatrick [6] proposed a model for spin glasses (the SK 
model) in which all Ising spins are coupled with each other through 
interactions which are given by independent Gaussian random numbers. 
The model was introduced to construct the mean-field theory of spin glasses. 
Making use of the replica method, they obtained various properties of 
spin glasses. Although the original SK solution involves a difficulty to yield 
negative entropy at low temperatures, the model is now resolved by the 
replica-symmetry-breaking solution due to Parisi [7]. 

The replica method is successful, but it is rather abstract since, by this 
method, the average over samples is carried out before examining 
thermodynamic properties of an individual sample.  In order to get 
more direct physical insights of the SK model, Thouless, Anderson and Palmer 
(TAP) [8] developed the mean-field theory in the phase space, by which one 
first treats an individual sample and then takes the average over samples.  
They proposed the free-energy form which contains the effect of the 
2-spin cluster besides the terms given by the conventional mean-field theory. 
The TAP free energy, properly derived afterwards [9,10],  well works to 
further clarify various features of spin glasses such as the marginal 
stability of the SG phase [11], the many-valley structure in the 
free-energy landscape [12], the number of local 
free-energy minima [13] and 
so on. It is now known that the TAP free-energy approach and the replica 
method are consistent with each other and provide complementary 
understandings of spin glasses [14,15].

The present work is motivated to develop such a TAP-like approach 
to the Hopfield model which is expected to play roles complementary to the 
AGS replica theory. Such an approach has been already described by 
M\'ezard, Parisi and Virasoro (MPV) in their text book [14].  
Based on the cavity method, which they have successfully developed to derive 
the TAP equations of states for the SK model, they have proposed the 
corresponding equations of states for the Hopfield model. 
We consider, however, that a part of their derivation has remained to be 
justified. 

The main purpose of the present paper is to derive the TAP free-energy 
expression for the Hopfield model directly by following the method due to 
Plefka [10], who derived the TAP free energy of the SK model. 
A crucial difference between the two models is that there exist correlations 
between different elements of the interaction matrix in 
the Hopfield model [14,16], while they are  not found in the SK model. 
Consequently the TAP free energy of the former consists of an infinite series 
of terms exhibiting such correlation (cluster) effects. 
Based on the TAP free energy derived, we analyze mostly nature of the SG phase 
of the model and compare the results with those obtained by the replica method 
as well as by the cavity method.  The derived TAP free energy is valid also in 
the retrieval FM phase, but the solution in this phase is left for a 
future study. 

In the next section we present the derivation of the TAP free energy of the 
Hopfield model. The PM-SG transition temperature $T_{\rm SG}$ is calculated in 
section 3. Section 4 is devoted to some related discussions including 
comparisons of the present results with those obtained by AGS and MPV. 

\vspace*{-3mm} 
\section{Derivation of the TAP Free Energy}

Our starting Hamiltonian is
\begin{eqnarray}
 H = - \sum_{ \langle i,j \rangle }J_{ij}S_{i}S_{j},
\end{eqnarray}
where $i(=1,2, \cdots ,N)$ denote spin (neuron) sites, and $S_i$ stand for 
spins (neurons) 
and take the values $\pm1$; the value $+1$ and $-1$ correspond to the neuron
which is firing and is not firing, respectively.  The summation is taken over
all spin (neuron) pairs.

The interaction (synaptic efficiencies) $J_{ij}$ are given by
\begin{eqnarray}
   J_{ij}=\left\{
   \begin{array}{ll}
   \displaystyle \frac{1}{N} \sum_{\mu=1}^{p}\xi_i^{\mu}\xi_j^{\mu} & \makebox[3em]{\rm for} i\neq j \\
   0 & \makebox[3em]{\rm for} i=j,
\end{array}
\right.
\end{eqnarray}
where $\xi_i^{\mu}$ take $\pm1$ and $\{\xi_i^{\mu}\}$ represent the $\mu$-th 
stored pattern.  Here we consider that $\xi_i^{\mu}$ are quenched, 
independent and random variables.  This means that $J_{ij}$ are also random 
variables.  One sees that $J_{ij}$ obey the Gaussian distribution with 
$\overline{J_{ij}}=0$ and $\overline{J_{ij}^2}=p/N^2$, where the overline 
indicates the average over samples (different 
realizations of $\{ J_{ij} \}$, or $\{\xi_i^{\mu}\}$'s).  
It should be noticed here that $\{ J_{ij} \}$ are 
not independent with each other, but have correlations between different 
$J_{ij}$'s [14,16]; for example, we see
\begin{equation}
\overline{J_{ij}J_{jk}J_{ki}}=\frac{p}{N^3}=\frac{\alpha}{N^2}.
\end{equation}
These non-zero correlations bring about new terms in the 
free energy (see below).

In order to obtain the free energy, we follow Plefka [10].
Introducing external fields $h_i^{\rm ex}$, we consider
\begin{equation}
\tilde{H}=aH-\sum_i h_i^{\rm ex} S_i.
\end{equation}
Then, we make the Legendre transformation to get the free energy as a
function of $m_i$,
\begin{equation}
F=-T \ln {\rm Tr} \ e^{-\beta \tilde{H}} + \sum_i h_i^{\rm ex} m_i.
\end{equation}
Here $T$ is the temperature ($\beta=1/T$, with $k_{\rm B}=1$) and 
$m_i = \langle S_i \rangle_a$, where $\langle \cdots \rangle_a$ denotes the 
expectation value with respect to $\tilde{H}$.  
We expand (5) with respect to $a$, i.e., 
\begin{eqnarray}
F(a)=\left. \sum_{n=0} \frac{1}{n!}\frac{\partial^n F}{\partial a^n}\right|_{a=0} a^n,
\end{eqnarray}
and then we put $a=1$.  Plefka showed
\begin{eqnarray}
\frac{\partial F}{\partial a} &=&  \langle H \rangle_a,  \\
\frac{\partial^2 F}{\partial a^2} &=& -\beta \langle H ( H - \langle H \rangle_a - \Lambda_1 ) \rangle_a, 
\end{eqnarray}
and obtained
\begin{eqnarray}
\left. \frac{\partial F}{\partial a}\right|_{a=0} &=& - \sum_{\langle i,j \rangle}J_{ij} m_im_j, \\
\left. \frac{\partial^2 F}{\partial a^2}\right|_{a=0} &=& - \beta \sum_{\langle i,j \rangle}J_{ij}^2 (1-m_i^2)(1-m_j^2),
\end{eqnarray}
where we have introduced
\begin{eqnarray}
\Lambda_n&=&\sum_i \frac{\partial^n h_i^{\rm ex}}{\partial a^n}(S_i-m_i) \nonumber \\
&=& \sum_i \frac{\partial}{\partial m_i} \left( \frac{\partial^n F}{\partial a^n} \right) (S_i-m_i).
\end{eqnarray}

Now we extend the calculation up to the 4-th order.  This calculation is 
fairly lengthy; we have made use of the algebraic programming system 
REDUCE-2.  The results thus obtained are as follows:
\begin{eqnarray}
\frac{\partial^3 F}{\partial a ^3} &=& \beta \langle H \rangle_a \frac{\partial \langle H 
\rangle_a }{\partial a} + \beta \langle H \Lambda_2 \rangle_a 
 + \beta^2 \langle H ( H - \langle H \rangle_a - \Lambda_1 )^2 \rangle_a, \\
\frac{\partial^4 F}{\partial a ^4} &=&
 3\beta \left( \frac{\partial \langle H \rangle_a }{\partial a} \right) ^2
+ \beta \langle H \rangle_a \frac{\partial^2 \langle H \rangle_a }{\partial a^2} + \beta \langle H \Lambda_3 \rangle_a \nonumber \\
& & - 3\beta^2 \langle H \Lambda_2 ( H - \langle H \rangle_a - \Lambda_1 ) \rangle_a 
- \beta^3 \langle H ( H - \langle H \rangle_a - \Lambda_1 )^3 \rangle_a ,
\end{eqnarray}
and
\begin{eqnarray}
\left. \frac{\partial^3 F}{\partial a ^3}\right|_{a=0} &=& 
-  4\beta^2 \sum_{\langle i,j \rangle}J_{ij}^3 m_im_j(1-m_i^2)
(1-m_j^2) \nonumber \\
& & - 6\beta^2  \sum_{\langle i,j,k \rangle}J_{ij}J_{jk}J_{ki}(1-m_i^2)(1-m_j^2)
(1-m_k^2),  \\
\left. \frac{\partial^4 F}{\partial a ^4}\right|_{a=0} &=&
- 2\beta^3 \sum_{\langle i,j \rangle}J_{ij}^4(15m_i^2m_j^2-3m_i^2-3m_j^2-1)  \nonumber \\
& & - 48\beta^3 \sum_{\langle i,j,k \rangle}J_{ij}J_{jk}J_{ki}(1-m_i^2)(1-m_j^2)(1-m_k^2) \nonumber \\
& & \makebox[5em]{} \times (J_{ij}m_im_j+J_{jk}m_jm_k+J_{ki}m_km_i) \nonumber \\
& & - 24\beta^3 \sum_{\langle i,j,k,\ell \rangle}J_{ij}J_{jk}J_{k \ell}J_{\ell i}(1-m_i^2)(1-m_j^2)(1-m_k^2)(1-m_\ell^2).
\end{eqnarray}
In the above, ${\langle i,j,k \rangle}$ and ${\langle i,j,k,\ell \rangle}$
denote that the summation should be taken over {\it inequivalent}
3-spin clusters and 4-spin clusters, respectively.

The free energy should be of the order of $N$, and therefore we have only to
pick up terms proportional to $N$ in  (9), (10), (14) and (15).
For the ferromagnetic Weiss model with $J_{ij}=1/N$, we can see that only 
(9) gives the contribution proportional to $N$, as it should.
In the SK model, the interactions $\{J_{ij}\}$ obey the simple Gaussian 
distribution with $\overline{ J_{ij} }=0$ and $\overline{ J_{ij}^2 }=O(1/N)$, 
and there is no correlation between different $J_{ij}$'s. Therefore, as TAP 
pointed out, equations (9) and (10) give the contribution of the order of $N$.
In the Hopfield model of present interest, equations (9) and (10) are of the 
order of $N$ as in the SK model. As mentioned in section 1, however, there 
exist correlations between different $J_{ij}$'s. This provides new terms 
to the free energy. To show this, we take the last term of 
 (14), as an example.  Its order of magnitude is estimated as 
\begin{eqnarray}
& &\sum_{\langle i,j,k \rangle}J_{ij}J_{jk}J_{ki}(1-m_i^2)(1-m_j^2)(1-m_k^2) \nonumber \\
&\sim& \frac{N(N-1)(N-2)}{6} \cdot \overline{J_{ij}J_{jk}J_{ki}} \nonumber \\
&\sim& \alpha N.
\end{eqnarray}
Similarly one can see that the last term of  (15) yields the contribution
of $O(N)$.  As for the other terms in  (14) and (15), one can see that
they can be neglected in the limit $N \rightarrow \infty$.
These analyses imply that 
$\left. \partial^n F / \partial a^n \right|_{a=0}$ for $n \geq 5$ also
provide the terms of $O(N)$, which are written in the form,
\begin{equation}
-n!\beta^{n-1} \sum_{\langle i_1,i_2,\cdots,i_n \rangle}J_{i_1i_2}J_{i_2i_3}\cdots J_{i_ni_1}(1-m_{i_1}^2)(1-m_{i_2}^2)\cdots (1-m_{i_n}^2).
\end{equation}
Their explicit derivation is given in Appendix A.

As a result, we have the following free energy,
\begin{eqnarray}
F = F_0 + F_{\rm cluster},
\end{eqnarray}
with
\begin{eqnarray}
F_0&=& - \sum_{\langle i,j \rangle}J_{ij}m_im_j 
 +T \sum_i \left( \frac{1+m_i}{2}\ln\frac{1+m_i}{2}
                + \frac{1-m_i}{2}\ln\frac{1-m_i}{2} \right), \\
F_{\rm cluster} &=& - \frac{1}{2} \beta \sum_{\langle i,j \rangle}
                 J_{ij}^2(1-m_i^2)(1-m_j^2) \nonumber \\
& &-\sum_{n=3}^{\infty}\beta^{n-1} \sum_{\langle i_1,i_2,\cdots,i_n \rangle}
                J_{i_1i_2}J_{i_2i_3}\cdots J_{i_ni_1} \nonumber \\
& & \makebox[8em]{} \times (1-m_{i_1}^2)(1-m_{i_2}^2) \cdots(1-m_{i_n}^2),
\end{eqnarray}
where the second term in  (19) is the entropy, which comes from $F(0)$ in (6).
The TAP equations of states described in terms of $\{m_i\}$ are determined by 
$ \partial F/\partial m_i =0$, i.e.,
\begin{eqnarray}
T \tanh ^{-1} m_i &=& \sum_j J_{ij}m_j -  \beta \sum_j
                 J_{ij}^2(1-m_j^2)m_i \nonumber \\
& &-2 \sum_{n=3}^{\infty}\beta^{n-1} \sum_{\langle i|j_1,j_2,\cdots,j_{n-1} \rangle}
                J_{ij_1}J_{j_1j_2}\cdots J_{j_{n-1}i} \nonumber \\
& & \makebox[5em]{} \times (1-m_{j_1}^2)(1-m_{j_2}^2) 
\cdots(1-m_{j_{n-1}}^2)m_i,
\end{eqnarray}
for $i=1,2,\cdots,N$, where $\langle i| j_1,j_2,\cdots,j_{n-1} \rangle$ means 
that the summation should be taken over {\it inequivalent} $n$-spin clusters 
with fixed $i$. With the substitution of $m_k^2$ appearing explicitly 
in (21) by the spin-glass order parameter $q=N^{-1}\sum_im_i^2$, equation (21) 
is rewritten as 
\begin{equation}
T \tanh ^{-1} m_i = \sum_j J_{ij}m_j  
- {\alpha \beta(1-q) \over 1-\beta(1-q)}m_i. 
\end{equation}
In deriving (22) we have used
\begin{eqnarray}
\sum_{\langle i|j_1,j_2,\cdots,j_{n-1} \rangle}
J_{ij_1}J_{j_1j_2}\cdots J_{j_{n-1}i}&=&
 \frac{(N-1)(N-2) \cdots (N-n+1)}{2} \cdot \frac{\alpha}{N^{n-1}} \nonumber \\
& \cong & \frac{1}{2}\alpha,
\end{eqnarray}
in which the factor 2 has been introduced, because we have
\begin{eqnarray}
J_{ij_1}J_{j_1j_2}\cdots J_{j_{n-1}i} = 
J_{ij_{n-1}} \cdots J_{j_2j_1} J_{j_1 i}.
\end{eqnarray}

\vspace*{-3mm} 
\section{Spin-Glass Transition Temperature}

Let us calculate the transition temperature, $T_{\rm SG}$, which separates the 
normal (disordered) and spin-glass phases. To do so, we expand  (22) up 
to the first order of $m_i$ and obtain 
\begin{eqnarray}
Tm_i = \sum_j J_{ij}m_j - \frac{\alpha}{T-1} m_i.
\end{eqnarray}
This implies that $T_{\rm SG}$ is given by the equation,
\begin{eqnarray}
T_{\rm SG} +\frac{\alpha}{T_{\rm SG}-1} - J_{\rm max}=0,
\end{eqnarray}
where $J_{\rm max}$ is the maximum eigenvalue of the interaction matrix 
${\hat J}$. It should be mentioned here that the condition
\begin{eqnarray}
J_{\rm max} \geq 1 + 2 \sqrt{\alpha}
\end{eqnarray}
should be satisfied to have real $T_{\rm SG}$.

Our task is then to calculate $J_{\rm max}$. In Appendix B, it is shown that
the distribution function of eigenvalues of $\hat{J}$ is given as follows:
\begin{eqnarray}
   \rho (\lambda)=\left\{
   \begin{array}{ll}
   \displaystyle \rho_0 (\lambda)+(1-\alpha)\delta(\lambda + \alpha) & \makebox[3em]{\rm for} \alpha \leq 1 \\
   \rho_0 (\lambda) &  \makebox[3em]{\rm for} \alpha>1,
\end{array}
\right.
\end{eqnarray}
with
\begin{eqnarray}
\rho_0 (\lambda) = \frac{1}{2\pi} \cdot \frac{\sqrt{(\lambda-1+2\sqrt{\alpha})
(1+2\sqrt{\alpha}-\lambda)}}{\lambda + \alpha},
\end{eqnarray}
where $\lambda$ stands for eigenvalues of $\hat{J}$. 
In Fig. 1 the behavior of $\rho_0 (\lambda)$, a continuous part 
of $\rho (\lambda)$, is shown for 
some $\alpha$. One notices at once that 
$\rho(\lambda)$ exhibits a quite different behavior from that of the 
independent Gaussian random matrix, for which it obeys the semi-circular 
law [11].  This is again the consequence 
of the non-zero correlations between the different matrix elements.
For $\alpha<1$ $\rho(\lambda)$ consists of a delta peak at $\lambda=-\alpha$ 
(whose amplitude is $1-\alpha$) and the continuous distribution 
$\rho_0(\lambda)$ around $\lambda=1$, i.e., 
$1-2\sqrt{\alpha} \leq \lambda \leq 1+2\sqrt{\alpha}$ (whose integrated 
amplitude is $\alpha$). Note that $\rho(\lambda)$ is normalized as 
$\int\rho(\lambda)d\lambda=1$. At $\alpha=1$ the delta peak merges to 
$\rho_0(\lambda)$, and for $\alpha>1$ $\rho(\lambda)$ exhibits a single
and broad peak.  As for the shape of $\rho_0(\lambda)$, we see from (29) 
that it becomes semi-circular and semi-elliptic for small and large 
$\alpha$, respectively. 

In any $\alpha$ the largest eigenvalue is given 
by the upper edge of $\rho_0(\lambda)$; $J_{\rm max}=1+2\sqrt{\alpha}$.
Then we rewrite  (26) to have
\begin{eqnarray}
\frac{(T_{\rm SG} -1-\sqrt{\alpha})^2}{T_{\rm SG} -1}=0.
\end{eqnarray}
This leads to $T_{\rm SG} = 1+\sqrt{\alpha}$. It is noted that $J_{\rm max}$ 
thus obtained is just on the boundary of the condition (27), or 
$T_{\rm SG}$ is given as a double root of (26). These circumstances are 
the same as those of the spin-glass transition temperature extracted 
by the TAP equation in the SK model.

\vspace*{-3mm} 
\section{Discussion}

The spin-glass transition temperature $T_{\rm SG}$ obtained by (30) 
coincides with the AGS result derived by the replica method. A further 
interesting comparison with the AGS result is on the expression of the 
entropy. To show this, let us rewrite $F_{\rm cluster}$ of (20) in terms 
of the spin-glass order parameter $q$ as we have done to derive (22). 
We obtain
\begin{eqnarray}
F_{\rm cluster}=\frac{1}{2} \alpha N \left\{ 1-q + T \ \ln [1-\beta (1-q)] 
\ \right\}.
\end{eqnarray}
The entropy coming from $F_{\rm cluster}$ is then given by 
\begin{eqnarray}
S_{\rm cluster}&=& - \frac{\partial F_{\rm cluster}}{\partial T} \nonumber \\
& = & -\frac{1}{2} \alpha N \left\{ \ln [1-\beta (1-q)] + \frac{\beta (1-q)}
{1-\beta (1-q)} \right\}.
\end{eqnarray}
This is exactly the same expression as that of the entropy 
in the limit $T\rightarrow 0$ calculated by AGS ($S_0=-(\partial F_0/
\partial T)=0$ in this limit). 
It becomes negative when it is evaluated in terms of the 
replica-symmetric solutions [5]. An expected proper solution is, as TAP argues 
for the SK model [8], that $(1-q)$ should vanish faster than 
$T$ as $T \rightarrow 0$ because we should have $S_{\rm cluster}=0$ at $T=0$.

In relation with the present result that $T_{\rm SG}$ is determined as a 
double root of (26), let us consider the susceptibility matrix 
$\chi_{ij}=\partial m_i /\partial h_j$.  It is known that ${\hat \chi}$ is 
given by $\hat{\chi}=\beta \hat{A}^{-1}$, where $\hat{A}$ is the Hessian 
matrix defined by $A_{ij}=\partial^2 (\beta F) / \partial m_i \partial m_j$.
Then we see that $\chi_{\rm max}$ diverges at $T_{\rm SG}$ as 
$\chi_{\rm max} \simeq (T- T_{\rm SG})^{-2}$, where $\chi_{\rm max}$ is the 
susceptibility of the eigenmode with the largest eigenvalue 
$J_{\rm max}$. The spin-glass susceptibility defined by $\chi_{\rm SG}=(1/N) 
\ {\rm Tr}\ \hat{\chi}^2$ is calculated as
\begin{eqnarray}
\chi_{\rm SG}= \int {\rm d}\lambda \frac{\rho(\lambda)}
{\left(T+ \alpha/(T-1)-\lambda \right)^2}
\end{eqnarray}
in the PM phase. Since $\rho(\lambda) \sim (1+2\sqrt{\alpha}-\lambda)^{1/2}$ 
near its upper edge, we obtain $\chi_{\rm SG} \sim (T - T_{\rm SG})^{-1}$. 
The replica method can provide the same result. These results 
described here indicate that nature of the PM-SG transition in the Hopfield 
model, including that the replica-symmetry-breaking takes place in the SG 
phase [5], is almost identical to that in the SK model.

The TAP equations of state for the Hopfield model was already discussed by 
MPV [14].  They made use of the cavity method twice. In the first step, 
one spin is added to the $N$-spin system, and the relations between quantities 
such as the free energy and the density of states of the $N$- and $(N+1)$-spin 
systems are examined to determine the distribution of field to the added
spin. Then the following TAP equations are derived
\begin{equation}
 m_i = \tanh \beta \left[ \sum_j J_{ij}m_j -  \beta(r_2-r_1)m_i \right],
\end{equation}
where $r_2 - r_1 = N^{-1}\sum_{\mu=1}^p (\langle \eta_\mu^2 \rangle - \langle \eta_\mu \rangle^2)$ with 
$\eta_\mu=N^{-1/2}\sum_{i=1}^N\xi_i^\mu S_i$. For the SK model this step 
alone gives rise to the TAP equations of interest [14].  For the Hopfield 
model, on the other hand, MPV introduced another `cavity method', in which 
the relevant relations are those of
quantities in the systems where $p$ and $(p+1)$ patterns are 
stored. This yields, for the replica-symmetric solution, 
\begin{equation}
 r_2 - r_1 = {\alpha \over \beta[1-\beta(1-q)]}.
\end{equation}
However equation (34) with (35) substituted does not coincide with our result, 
equation (22). Since the factor $\beta(1-q)$ in the numerator of the second 
term 
of (22) is missing, the MPV equations do not reproduce the proper 
$T_{\rm SG}$. We suppose that the origin of the discrepancy would lie in the 
second step of the cavity method in the MPV argument. 

Finally we make a comment on the work by Geszti [2].  
Starting from the equations $m_i= \tanh (\beta \sum J_{ij}m_j)$, he 
derived a set of the self-consistent equations for the retrieval FM order 
parameter $m$, the random overlap parameter $r$, and $q$, which 
coincides with those due to AGS derived by the replica theory.
In his heuristic argument, however, the terms in (21) coming from 
$F_{\rm cluster}$ of (20) are ignored. His argument is similar to 
the one by which the self-consistent equation for $q$ 
of the SK model is derived, and which is criticized in [15].
A proper solution of (21) in the retrieval FM phase is our next 
concern.

To conclude we have developed a TAP-like mean-field theory on the Hopfield 
model, by which we first analyze thermodynamics of individual sample with 
fixed $\{J_{ij}\}$, or $\{ \xi_i^\mu \}$'s and then take the average over 
samples. In contrast to the SK model for spin glasses where only the 2-spin 
cluster effect is vital, it has been shown that a series of clusters, 
composing a large number of spins, play an important role in the Hopfield 
model. This gives rise to the TAP free energy which contains an infinite 
number of terms. Based on it we have investigated the PM-SG transition in 
the Hopfield model to find that its nature is almost identical to that in 
the SK model. We consider that the present TAP free-energy approach is 
useful in studying neural networks of a mean-field type since it will 
provide us complementary information to the replica method.

\newpage
\setcounter{equation}{0}
\renewcommand{\theequation}{A\arabic{equation}}
\noindent
\section*{ Appendix A. Derivation of equation (17)}

Here we discuss $\partial^n F / \partial a^n$, from which we have terms
of the order of $N$.  We notice that the last terms of  (14) and (15) come 
from the last terms of  (12) and (13), respectively.
Therefore we concentrate, in $\partial^n F / \partial a^n$, on the term
\begin{eqnarray}
(- \beta)^{n-1} \langle H ( H - \langle H \rangle_a - \Lambda_1 )^{n-1} \rangle_a.
\end{eqnarray}
It is easily seen that $\partial^n F / \partial a^n$ contains the above term if one notices
\begin{eqnarray}
\frac{\partial \langle R \rangle_a}{\partial a} =
\left\langle \frac{\partial R}{\partial a} \right\rangle_a
- \beta \langle R ( H - \langle H \rangle_a - \Lambda_1 ) \rangle_a.
\end{eqnarray}
On the other hand, we have
\begin{eqnarray}
H - \langle H \rangle_a - \Lambda_1 = - \sum_{\langle i,j \rangle}J_{ij}(S_i-m_i)(S_j-m_j),
\end{eqnarray}
and therefore  (A1) can be written by
\begin{eqnarray}
-\beta^{n-1} \left\langle \sum_{\langle i,j \rangle}J_{ij}S_iS_j
 \left[ \sum_{\langle i,j \rangle}J_{ij}(S_i-m_i)(S_j-m_j) \right]^{n-1}
\right\rangle_a.
\end{eqnarray}
This provides equation (17) in the text, together with other irrelevant terms.

\vspace{15mm}
\setcounter{equation}{0}
\renewcommand{\theequation}{B\arabic{equation}}
\noindent
\section*{ Appendix B.  Eigenvalue Distribution of $\hat{J}$}
\par
Following Bray and Moore [11], we write down the distribution function of 
eigenvalues of $\hat{J}$ as 
\begin{eqnarray}
\rho(\lambda) &=& \frac{1}{N} \sum_i \delta (\lambda - \lambda_i) \nonumber \\
&=& \frac{1}{\pi} {\rm Im} \left[ \frac{1}{N} \sum_i G_{ii}(\lambda - {\rm i} \epsilon) \right],
\end{eqnarray}
where $\epsilon$ is a positive infinitesimal and $G_{ii}$ are the diagonal 
elements of the matrix Green function
\begin{eqnarray}
\hat{G}(\lambda) = (\lambda \cdot \hat{1} - \hat{J})^{-1},
\end{eqnarray}
with $\hat{1}$ being the unit matrix.
Then we make use of the so-called locator expansion to have
\begin{eqnarray}
G_{ii} &=& \frac{1}{\lambda}+\frac{1}{\lambda} \sum_j \left( J_{ij} \frac{1}{\lambda}J_{ji} \right) \frac{1}{\lambda} 
 + \frac{1}{\lambda} \sum_{j,k} \left( J_{ij} \frac{1}{\lambda} J_{jk} \frac{1}{\lambda} J_{ki} \right) \frac{1}{\lambda} + \cdots \nonumber \\
&=& \frac{1}{\lambda} + \Delta + \lambda \Delta^2 + \lambda^2 \Delta^3 + \cdots \nonumber \\
&=& \frac{1}{\lambda (1- \lambda \Delta)}.
\end{eqnarray}
Here $\Delta$ consits of an infinite series of terms due to the existence 
of correlations between different matrix elements of $\hat{J}$ (see section 2).
Indeed, it is given by
\begin{eqnarray}
\Delta &=& \frac{1}{\lambda} \sum_j \left( J_{ij} \overline{G} J_{ji} \right) \frac{1}{\lambda} 
 + \frac{1}{\lambda} \sum_{(i|j,k)} \left( J_{ij} \overline{G} J_{jk} \overline{G} J_{ki} \right) \frac{1}{\lambda} \nonumber \\
& & + \frac{1}{\lambda} \sum_{(i|j,k,\ell)} \left( J_{ij} \overline{G} J_{jk} \overline{G} J_{k \ell} \overline{G} J_{\ell i} \right) \frac{1}{\lambda} + \cdots \nonumber \\
&=& \frac{1}{\lambda^2} \sum_{n=2}^{\infty} \overline{G}^{n-1}
\sum_{(i_1 | i_2,i_3, \cdots ,i_n )}J_{i_1i_2}J_{i_2i_3}\cdots J_{i_ni_1},
\end{eqnarray}
where we have introduced $\overline{G}$ by
\begin{eqnarray}
\overline{G} = \frac{1}{N} \sum_i G_{ii}
\end{eqnarray}
to take into account the renormalization. In the above, 
$(i_1 | i_2,i_3, \cdots ,i_n)$ means that the summation should be taken over
$n$-body cluster for fixed $i_1$; we see
\begin{eqnarray}
& & \sum_{(i_1 | i_2,i_3, \cdots ,i_n )}J_{i_1i_2}J_{i_2i_3}\cdots J_{i_ni_1} 
\nonumber \\
&\simeq& (N-1)(N-2) \cdots (N-n+1) \overline{J_{i_1i_2}J_{i_2i_3}\cdots J_{i_ni_1}} \nonumber \\
&\simeq& \alpha.
\end{eqnarray}
Then we have
\begin{eqnarray}
\Delta = \frac{\alpha}{\lambda^2} \cdot \frac{ \overline{G}}{1-\overline{G}}.
\end{eqnarray}
From (B3), (B5) and (B7) we obtain 
\begin{eqnarray}
\overline{G}= \frac{1}{\lambda [ 1 - \alpha \overline{G} / \lambda (1-\overline{G})]},
\end{eqnarray}
which is solved as
\begin{eqnarray}
\overline{G}=\frac{1}{2(\lambda + \alpha)} \left[ \lambda + 1 \pm 
\sqrt{(\lambda + 1)^2 - 4(\lambda + \alpha)} \right].
\end{eqnarray}
The above solution yields the imaginary part of $\overline{G}$ as follows:
\begin{eqnarray}
{\rm Im} \overline{G} = \frac{\sqrt{4(\lambda + \alpha)-(\lambda + 1)^2}}{2(\lambda + \alpha)}+ \pi  C  \delta (\lambda + \alpha),
\end{eqnarray}
where
\begin{eqnarray}
   C=\left\{
   \begin{array}{ll}
   \displaystyle 1-\alpha &  \makebox[3em]{\rm for} \alpha \leq 1 \\
   0 & \makebox[3em]{\rm for} \alpha>1.
\end{array}
\right.
\end{eqnarray}
This result together with  (B1) and (B5) gives us equation (28) 
in the text.
\newpage
\noindent
\begin{center}
{\bf References}
\end{center}
\par
\parskip 0mm
\begin{enumerate}
\renewcommand{\labelenumi}{[\arabic{enumi}]}
\item Hopfield J J 1982 {\it Proc. Natl. Acad. Sci. USA} {\bf 79} 2554
\item Geszti T 1990 {\it Physical Models of Neural Networks} (Singapore: World Scientific)
\item Hertz J, Krogh A and Palmer R G 1991 {\it Introduction to the Theory of Neural Computation} (Tokyo: Addison-Wesley)
\item Peretto P 1992 {\it An Introduction to the Modeling of Neural Networks} (Cambridge: Cambridge Univ. Press)
\item Amit D J, Gutfreund H and Sompolinsky H 1985a {\it Phys. Rev.} {\bf A32} 1007; 1985b {\it Phys. Rev. Lett.} {\bf 55} 1530; 1987 {\it Ann. Phys.} {\bf 173} 30
\item Sherrington D and Kirkpatrick S 1975 {\it Phys. Rev. Lett.} {\bf 35} 1792
\item Parisi G 1979 {\it Phys. Lett.} {\bf 73A} 203; 1980 {\it J. Phys. A: Math. Gen.} {\bf 13} L115, 1101 and 1887; 1983 {\it Phys. Rev. Lett.} {\bf 50} 1946
\item Thouless D J, Anderson P W and Palmer R G 1977 {\it Phil. Mag.} {\bf 35}  593
\item Nakanishi K 1981 {\it Phys. Rev.} {\bf B23} 3514
\item Plefka T 1982 {\it J. Phys. A: Math. Gen.} {\bf 15} 1971
\item Bray A J and Moore M A 1979 {\it J. Phys. C: Solid State Phys.} {\bf 12} L441 
\item Nemoto K and Takayama H 1985 {\it J. Phys. C: Solid State Phys.} {\bf 18}  L529
\item Bray A J and Moore M A 1980 {\it J. Phys. C: Solid State Phys.} {\bf 13} L469
\item M\'ezard M,  Parisi G and  Virasoro M A 1987 {\it Spin Glass Theory and Beyond} (Singapore: World Scientific)
\item Fischer K H and Hertz J A 1991 {\it Spin Glasses} (Cambridge: Cambridge Univ. Press)
\item Kinzel W 1985 {\it Z. Phys.} {\bf B60} 205
\end{enumerate}

\vspace*{2cm}
\begin{center}
{\bf Figure Captions}
\end{center}
\begin{enumerate}
\renewcommand{\labelenumi}{Fig. \arabic{enumi}:}
\item The distribution function $\rho_0 (\lambda)$ for 
$\alpha=$0.1, 0.5, 1, 1.5 and 2.
\end{enumerate}
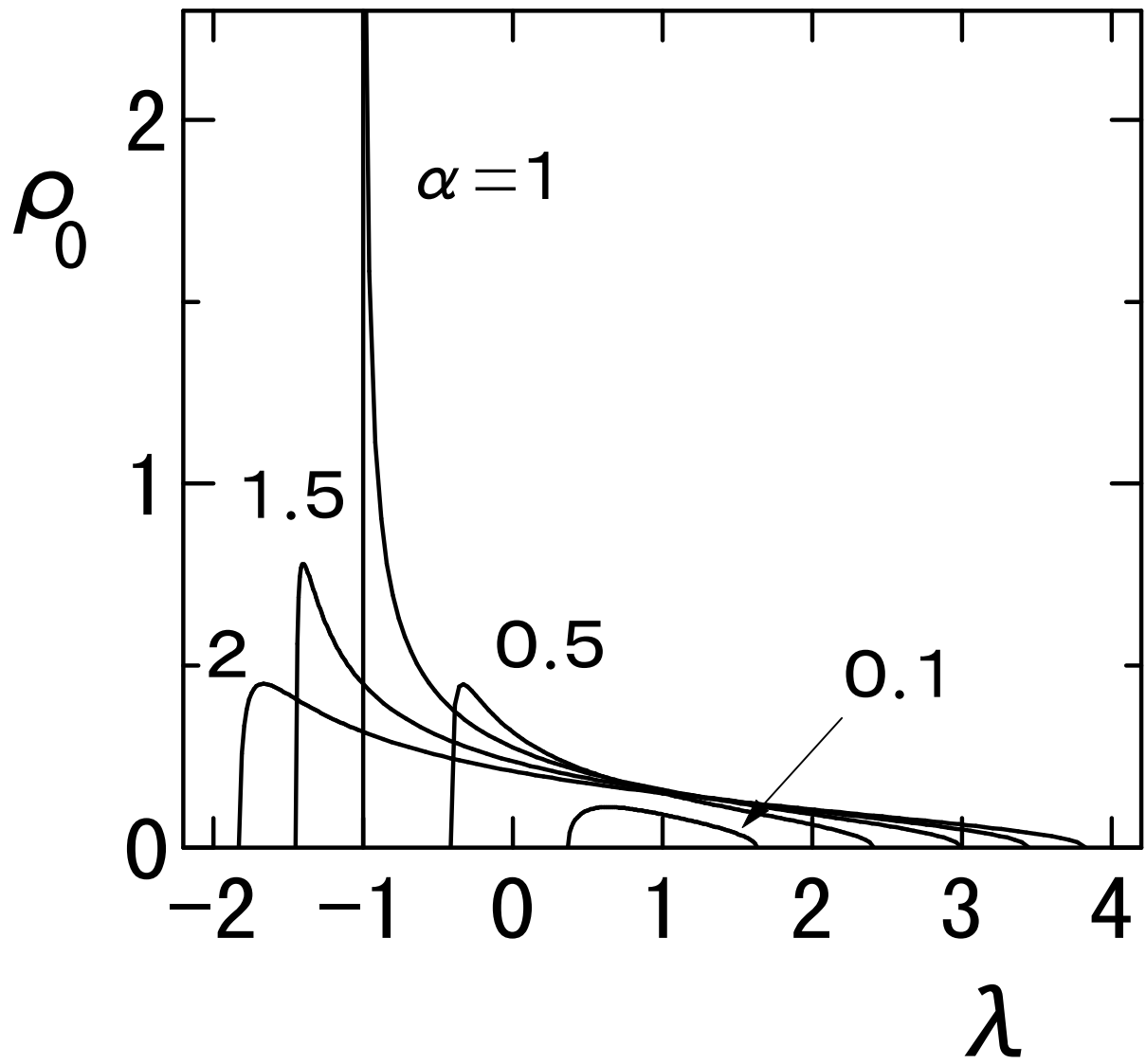
\end{document}